# A Novel Nano Tomography Setup for Material Science and Engineering Applications

Dominik Müller[1]*, Jonas Graetz[2], Andreas Balles[2], Simon Stier[3], Randolf Hanke[1,2] and Christian Fella[2]

1 Chair of X-Ray Microscopy, University of Würzburg, Germany

2 Nano-Tomography Group, Fraunhofer Development Center X-Ray Technology EZRT, Würzburg, Germany

3 Center Smart Materials and Adaptive Systems, Fraunhofer Institute for Silicate Research ISC, Würzburg, Germany

dominik.mueller@physik.uni-wuerzburg.de

**Abstract**

In a comprehensive study on several samples we demonstrate for our laboratory-based computed tomography system resolutions down to 150nm. The achieved resolution is validated by imaging common test structures in 2D and Fourier Shell Correlation of 3D volumes. As representative application examples from nowadays material research, we show metallization processes in multilayer integrated circuits, ageing in lithium battery electrodes, and volumetric of metallic sub-micrometer fillers of composites. Thus, our laboratory system provides the unique possibility to image non-destructively structures in the range of hundred nanometers, even for high density materials.

# 1 Introduction

Research and application of high resolution X-ray microscopy is still mainly carried out in large synchrotron facilities.[1,2] Nevertheless, especially optics-based full-field microscope setups have already made the leap into the laboratory[3], and such commercial devices can achieve resolutions below 100 nm.[4,5] Due to technical challenges in the fabrication of X-ray optical elements, these kinds of devices are restricted to low energies (mostly <9 keV), which strongly limits the possible applications in the field of material research.

A different approach for laboratory based nano computed tomography (CT) is downsizing the X-ray source of the projection magnification setup, already known from micro CT.[3,6] This can be done by using thin, specifically structured layers or nanometer-fine needle tips as electron beam targets. For this purpose, modified scanning electron microscopes are used in order to utilize the already existing prefocused electron beam and electron optics.[6–8] Also here resolutions below 100 nm have been demonstrated.[7] By avoiding X-ray optical components, these systems have less restrictions concerning the sample size, because the sample does not need to fit inside the optical depth of field, which simplifies the extension from radiography to computed tomography. However, a major drawback of these electron microscope based systems is that samples have to be transferred into vacuum close to the X-ray source, requiring vacuum resistance and conductivity of the objects.

In contrast, high-resolution X-ray sources equipped with thin transmission targets do not need a vacuum environment. Thus, with the recent availability of such a source,[9,10] it is now possible to build more versatile applicable computed tomography devices with resolutions of only a few hundred nanometers.[11,12]

In the following, we present a custom developed laboratory nano CT setup based on such a transmission nano focus source as a versatile tool for 3D imaging with resolutions down to 150 nm named "ntCT", designed to examine larger and higher absorbing objects than with the other aforementioned laboratory methods.

# 2 Introduction to the selected illustrative samples

We demonstrate the practical performance of our system by means of three exemplarily selected problems from the field of materials research. The first example is chosen from the field of integrated circuits. Due to the progressive miniaturization and more complex three-dimensional structures, new challenges for analysis methods are constantly arising.[13,14] As example to increase capacity and reduce bit costs of memory structures, multi-layer systems[15–17] as well as three-dimensional structures [18,19], were established. In order to connect the different planes of a multilayer memory structure in a microchip, so-called through silicon vias (TSV) are used. For this purpose, countless holes with high aspect ratios are etched into the wafer and then metallized with highly conductive tungsten or copper. With increasing number of layers, also the required aspect ratio increases, whose quality must be ensured allover the wafer. The critical characteristics of these contacts are dimension, depth and shape as well as insufficient metallization or enclosed voids.[14,20] It has already been successfully demonstrated that CT with resolutions down to 500 nm is a useful tool to characterize contacts with several micrometers in diameter.[20–22] However, current generation contacts are narrower than the resolution of these devices.[14,23] Therefore, nano CT methods are needed to meet the increased requirements for an analysis tool.

Another typical application example comes from the field of battery research.[24] Here, diverse research areas are interested in the microstructure of the electrodes. The electrode of a battery usually consists of a porous granular active phase system. Redox reactions in batteries during charge and discharge events generally imply phase transitions, lattice volume change, stress formation, grain boundary weakening and particle break-up. These processes occur on several length and time scales and are called chemo-mechanical interactions. Thus they contribute significantly to the complex fading mechanisms of battery electrodes. Since these are creeping processes, they do not occur simultaneously in the individual particles of the electrode. It is therefore particularly interesting to obtain an overview of the condition of several particles at a certain point in time, in order to draw conclusions about the ageing mechanism. These information serve also as a starting point for the simulation of new battery types.[25,26] The nano CT data has the potential to bridge the gap between experimentally determined macroscopic properties and simulation results.

The third application example in Figure 6 represents novel smart materials. Here, the research concentrates on creating a material composition with new properties by combining known materials in a special microstructure to provide electrical or magnetic properties, for example for sensors. The specimen of a smart material shown here is a highly electrically conductive inorganic polymer mixture, that retains its conductivity of about 100 S/cm not only under bending but also under high elongation.[27] It is therefore a promising candidate for wearable sensors. In addition to the degree of filling, the conductivity depends on the spatial and orientational distribution of the particles. Both parameters are initially defined to a large extent by the manufacturing process, for example the mixing or coating process for the liquid precursor, and are temporarily fixed after the elastomer has been crosslinked. However, subsequent deformation of the cured matrix inevitably leads to stretching or compression of the particle spatial distribution. As for all anisotropic particles, the deformation of the matrix also affects their orientation.[28,29] The actual spatial and orientational distribution of the particles are therefore strain-dependent and, taking viscoelastic effects into account, additionally time-dependent. Typically, the effects that occur are reversible, however, the particles are mobile in the matrix, as they are not covalently bound. Cyclic loading in particular can therefore lead to a permanent change in spatial and orientational distribution. For the optimization of the composites it is therefore of interest to record these parameters non-destructively in different strain and aging states. While destructive electron tomography has already been used to examine nanoparticle composites[30], nano CT is potentially more suitable for the analysis of microparticle systems due to the larger sample volume.

## 3 Experimental Setup

### 3.1 Device Introduction

The ntCT (Figure 1) follows the basic of projection magnification. It therefore consists of only a few core components mentioned below and does not require any further (X-ray) optical elements for imaging. The utilized X-ray source is the Excillum Nanotube N2 60 kV [10] with a 500 nm thick tungsten transmission target. According to the manufacturer's specifications, it achieves a 300 nm FWHM electron spot at 200 mW target power under high-resolution settings. This information is approximately consistent with our own investigations.[31] The source features an internal validation of the actual electron spot size as a measure of the achievable resolution.

For image acquisition we use a DECTRIS EIGER2 R[32] hybrid photon counting detector with a CdTe sensor. The detector has an active area consisting of 2070x514 square pixels, each 75 µm wide. It possesses two variable energy thresholds, so that only photons with an energy above the set level are counted simultaneously in two separate recordings. Based on the direct converting design, the detector has a high dynamic range due to the absent detector background and a high quantum efficiency of 0.9-1.0 (up to 25 keV), only limited by photons hitting edges and corners of pixels.[33,34] As shown in Figure 1 (b), the setup also has an optional DECTRIS detector with Si sensor. For all measurements shown in this article exclusively the aforementioned CdTe detector was used.

The exact positioning of the sample is realized by a 10-axes nano manipulator made of laboratory standard piezo stick-slip positioners (SmarAct GmbH) assembled according to our own design with 6 degrees of freedom (DOF), including a high-precision air-bearing rotary table (Leuven Air Bearings Nv) for CT application. A 3-axes manipulator made of linear spindle axes (HUBER Diffraktionstechnik GmbH & Co. KG) is used for placing the detector at distances between 250 mm to 650 mm from the source to set the magnification in a range from several micrometer down to 50 nm voxel sampling. Typical parameters for a CT scan are voxel samplings around 100 nm at 200 µm FOV resulting in about 500 counts/(s·pixel) at the detector. The entire system is packaged in temperature-controlled double-walled radiation protection enclosure. This cabin sits vibration damped on a massive granite.

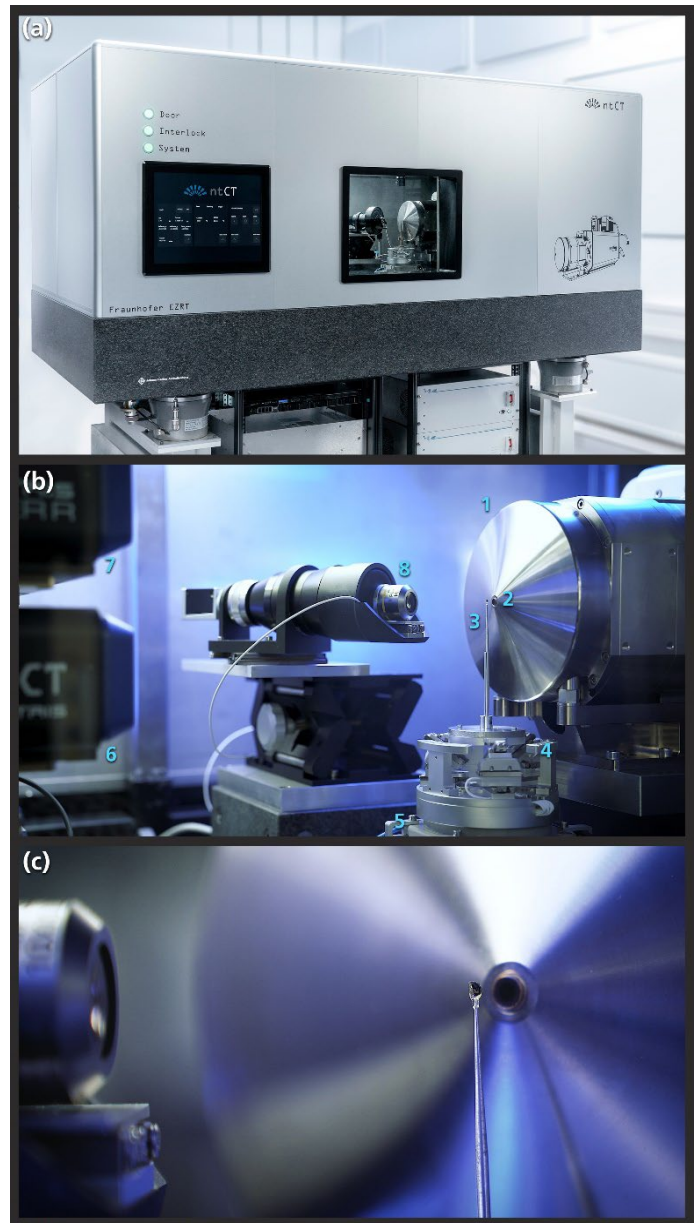

**Figure 1** Compact and integrated design of our laboratory nano CT setup. (a) Exterior view of the system with temperature-controlled lead booth and air-bearing granite for vibration damping. (b) Inside view into the device with 1 nanofocus X-ray source, 2 exit window and at the same time tungsten transmission target on a diamond substrate, 3 sample mounted on a needle-shaped holder, 4 6-DOF hexapod made of piezo linear axes for fine adjustment of the sample on the rotation axis, 5 air-bearing rotary stage (and in the picture not visible below three further linear axes for positioning of the sample holder), 6 DECTRIS EIGER2 R CdTe detector on 3 linear axes, 7 optional DECTRIS detector with SI sensor, 8 optical microscope to support the sample adjustment. (c) Detailed view of the sample position in front of the exit window of the X-ray source with mounted specimen.

### 3.2 Sample and measurement preparation

In best case, the diameter of the object is cut approximately to the desired field of view to avoid artifacts in the peripheral areas. The sample is then fixed on the tip of a needle-shaped sample holder, mounted on the manipulator and moved automatically to the examination position.

The reconstruction of the data is performed with an in-house developed filtered back-projection methodology and, depending on the sample, with a subsequent phase retrieval. In this step, the measured data is also corrected for deviations which inevitably occur in nano CT scans, such as sample drift on the sub-micrometer scale and irregularities in detector pixels sensitivity.

### 3.3 Validation of the resolution

To evaluate the 2D resolution of the system, a Siemens star test pattern was inspected. The sample was located 160 µm in front of the 100 µm thick exit window of the source, with the detector positioned at a distance of 300 mm. This leads to a sampling of 71 nm. The Siemens star consists of 1.5 µm lithographically etched tungsten made by ZonePlates Ltd. and has a smallest structure size of 150 nm half period in the center.

To assess the obtained resolution in 3D, the measurements of the three samples shown were evaluated using the Fourier Shell Correlation (FSC) method.[35][i] Here, the FSC is a measure of the normalized cross-correlation between the spatial frequencies of two equivalent 3D datasets in Fourier space. To create these two datasets, the individual tomographies were divided equally (even and odd projections) and reconstructed separately. The achieved resolution is then obtained after the FSC comparison of the two volumes when reaching the half-bit criterion corresponding to the sample.[36] For visual confirmation of the resolution, the contrast in presumably periodic structures of the semiconductor sample were analyzed and then compared to the previous results.

### 3.4 Objects and Parameters for the presented sample measurements

As the first sample we show exemplarily the internal structures of a commercially available semiconductor memory (SanDisk 32 GB micro SDHC UHS-I).[37,38] Second we display the active material phase $LiNiCoAlO_2$-$LiCoO_2$ of a commercial Pouch-Cell (Kokam High-Energy 560 mA).[39] Finally, we show a sample of a promising silicone-based conductive elastomer produced by Fraunhofer ISC Research Institute.[27] The material consists of a matrix of silicone elastomers and is filled with 20 vol% flake-shaped particles of silver-coated copper (Cu/Ag - 60/40). The highly anisotropic flakes have a thickness of approx. 100 nm, a diameter of approx. 4 µm and a high electrical conductivity. Table 1 summarizes the recording parameters used for the individual measurements.

| Sample | General Settings | | | | Detector | | Source | |
|---|---|---|---|---|---|---|---|---|
| | **SOD** | **SDD** | **Recordings** | **Sampling** | **Exposure Time** | **Threshold** | **Target Power** | **Voltage** |
| *Semiconductor* | *0.80 mm* | *575 mm* | *2800* | *98 nm* | *15 s* | *5 keV* | *165 mW* | *60 kV* |
| *Li-Battery* | *0.63 mm* | *425 mm* | *3600* | *105 nm* | *30 s* | *5 keV* | *67 mW* | *60 kV* |
| *Conductive Elastomer* | *0.46 mm* | *280 mm* | *2400* | *119 nm* | *10 s* | *8 keV* | *160 mW* | *60 kV* |

**Table 1** Overview of acquisition parameters for the shown computed tomography measurements with sample object distance (SOD), sample detector distance (SDD), number of exposures, effective sampling size, single image exposure time, lower detector threshold, power on the X-ray target and acceleration voltage.

### 4 Results

## 4.1 Resolution evaluation

Figure 2 shows the background corrected image of the center of the Siemens star with an exposure time of 300 s. In the enlarged section, even the innermost lines with the highest spatial resolution of 3300 lp/mm (corresponding to 150 nm half period) are clearly visible.

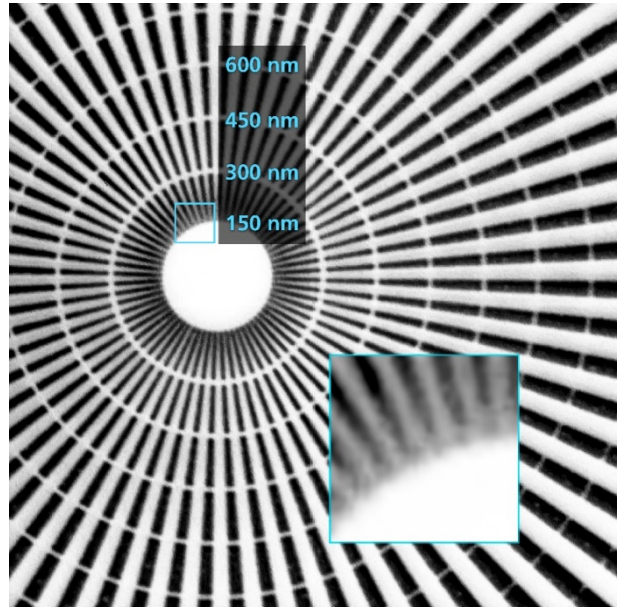

**Figure 2** Siemens star test pattern with 300 s exposure time. Inner Structure with 150 nm lines and spaces are clearly visible.

To characterize the systems modulation transfer function (MTF) as a more quantitative reference for the resolution, the contrast visibility in the Siemens star pattern can further be analyzed. In a previous study we have already shown that the MTF of the imaging system (Figure 3) can be approximately described by the superposition of a Gaussian PSF with 250 nm full width at half maximum (FWHM) with an additional background of about 1.1 µm FWHM. This finding could be attributed to the characteristics of the X-ray source by further evaluating the shape of the electron beam on the target. For this purpose, the focus of the electron beam was manually scanned over edges on the patterned transmission target, to obtain an independent measure of the source point spread function by evaluating the radiation intensity measured in this process. For a more detailed description of the method, please refer to the previous study indicated.[31] For the visible resolution, which corresponds to the 10% MTF level, 3200 lp/mm is achieved with the high resolution portion (narrow PSF). This is consistent with the visible 150 nm structures in the above shown Siemens star image.

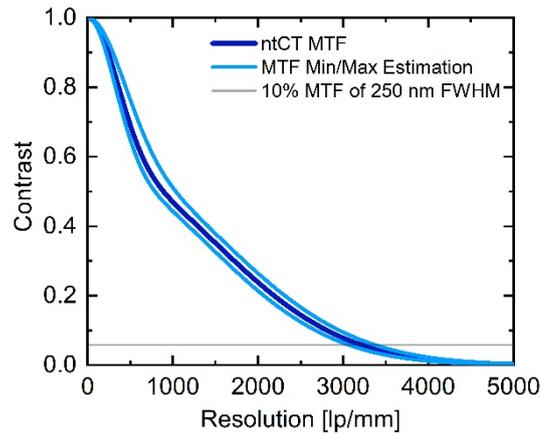

**Figure 3** 2D MTF graph derived from the Siemens star image. As the electron spot in the source can approximately be described by the superposition of a narrow and a wide focal spot, the MTF is also composed of a narrow 250 nm FWHM and a background share of 1.1 µm FWHM.[31][ii]

The 2D MTF defines the upper limit of the resolution achievable in tomographic imaging, which is subject to many other influencing factors such as the accuracy of the axes or drift in the individual components. To quantify the achieved resolution in 3D, the FSC was calculated from the volume data and plotted in Figure 4 together with the corresponding half-bit criterion.

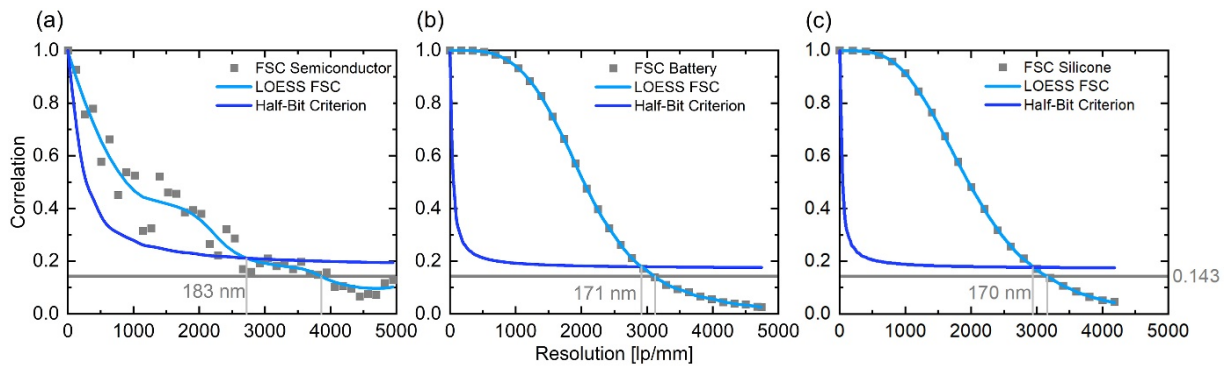

**Figure 4** FSC curves of the three samples shown below, (a) the semiconductor memory, (b) the lithium battery and (c) the conductive elastomer, and the corresponding half-bit criterion curves plotted versus spatial resolution. The FSC data were smoothed with a LOESS algorithm. The intersection points between both curves are shown in each case. Additionally, the less valid 1/7 level is plotted. The achieved resolutions are in the range of 170 nm - 185 nm and thus confirm the expectation slightly above the determined 2D resolution.

The achieved resolution is in the range of 170 nm - 185 nm for all three samples and therefore confirms the expectation. The result of the conductive elastomer has to be considered with reservation, since the determined 170 nm are close to the sampling rate. However, the value matches the other results.

As an illustrative example of the resolutions achieved, the local image contrast from the line profiles of the gray values over various structures marked in the section of the semiconductor scan were evaluated in Figure 5. In the adjacent plot, these points, which can be roughly assigned to the structure sizes 250 nm, 175 nm and 100 nm, were plotted against the MTF previously determined from the siemens star. For simplicity and since we do not have the design parameters, we have assumed these structures to be periodic and with regular duty cycle. This evaluation does not represent an analytical evaluation, but it can be shown that the expected behavior of the visibility of real structures in a sample can be confirmed.

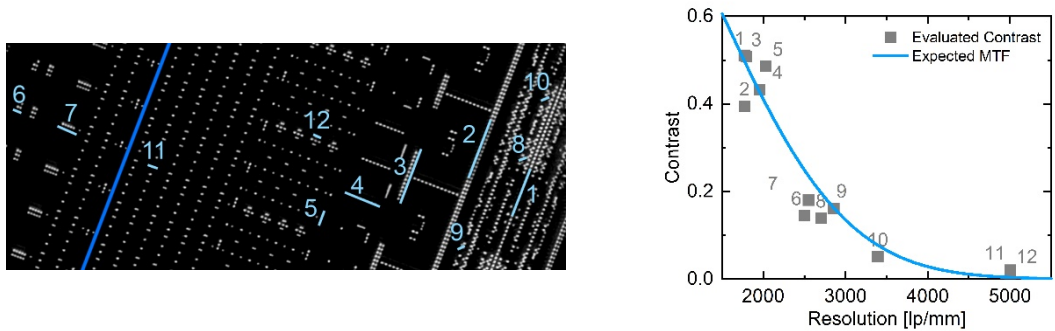

**Figure 5** Application of the local image contrast ratio at different structures in the horizontal section[iii] of the semiconductor memory data set which are assumed to be approximately periodic and comparison with the previously determined MTF. The corresponding areas in the image are marked accordingly. Points 11 and 12 are roughly on par with an expectable noise level and should therefore only be evaluated as image contrast with caution.

## 4.2 Semiconductor Sample

The reconstructed scan of the semiconductor memory chip is shown in Figure 6. The upper image a) shows a 3D rendering of the sample volume. In particular, the tower-like contacts made of tungsten can be seen, which connect the 48 storage layers to the electronics. b) shows a schematic enlargement of a single contact from the SI substrate. A cross-section of the volume and enlarged sections of it are shown in c-e. Based on publicly available information, these structures are expected to be between 100 - 200 nm,[15,16,23,40,41]. In the enlarged view e), two additional structure sizes were determined from the image and also specified. This fine, track-like structure consists of two rows of parallel plated-through contacts and has a period of 568 nm in length and 400 nm in lateral distance, corresponding to a frequency of 2500 lp/mm. Images f-i) show various vertical sections through the volume as shown in c). As expected from a commercial volume produced memory chip, the recognizable structures are uniform, parallel to each other and perpendicular to the other layers. There are also no other obvious defects visible in the metallization. Of particular interest is the visibility of the staircase structure typical for V-Nand memory[40–42] at the outer edge of the chip, as shown in f). With rising number of layers, the complexity of this staircase area increases, which is created in an elaborated multi-stage etching process. An exact positioning of the steps is necessary to contact the storage cells correctly.[42]

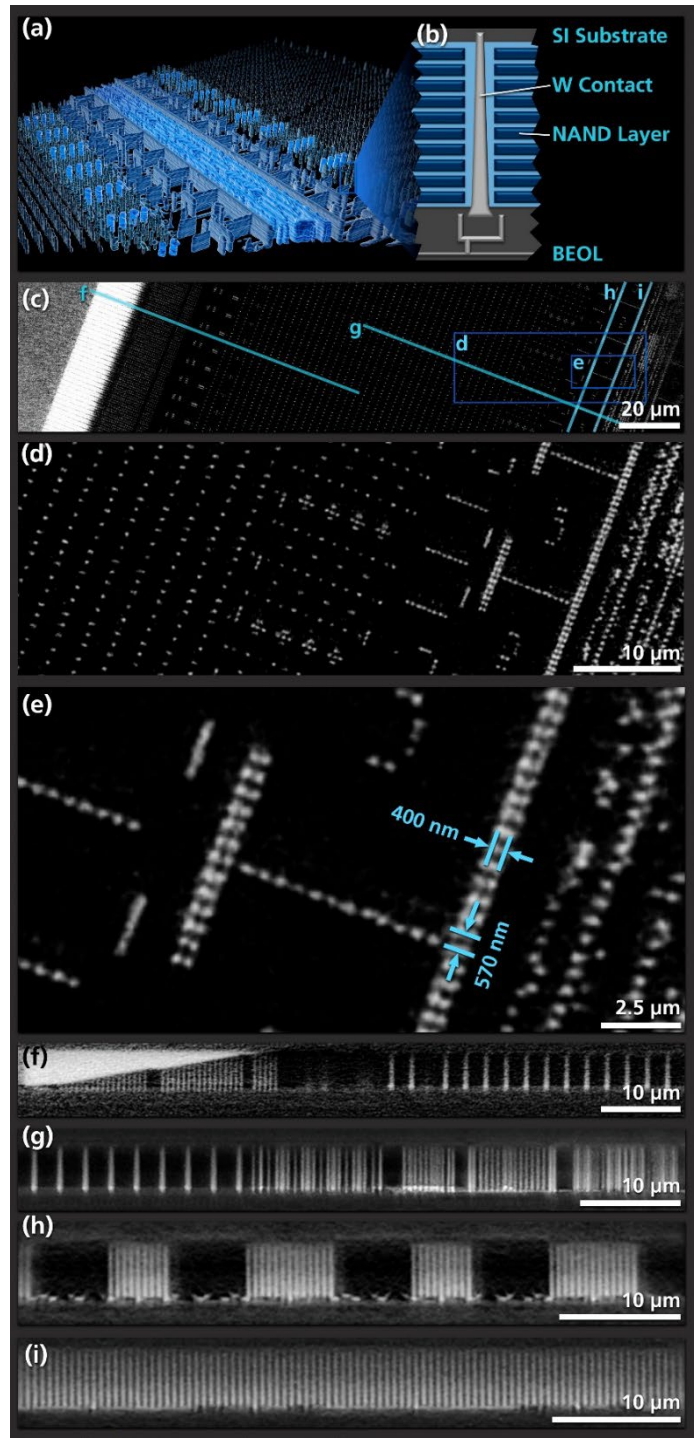

**Figure 6** Reconstructed nano CT scan of a semiconductor memory chip from a commercial micro sd card (SanDisk 32 GB microSDHC UHS-I) measured with a sampling of 98 nm. (a) a 3D rendering[iv] of the internal metallized structures, where the individual contacts with an expected size between 100 - 200 nm can be clearly identified. (b) schematic of a single contact pillar connecting the multiple wordlines. (c) volume cross-section from the reconstruction with memory array in the center and a section of the peripheral circuits to the right. (d) - (e) enlarged details of the sectional view with dimensions for specific features, (f) – (i) vertical sections through the volume corresponding to the markings in (c). (f) wordline staircase at the edge of the chip. (h) – (i) vertical view of the peripheral circuits.

## 4.3 Battery Research

Figure 7 shows the reconstruction of the Li battery electrode. The three-dimensional rendering in a) shows the shape and distribution of the active phases in the electrode. The corresponding false color assignment to the expected components is based on the gray value, shape and structure of the individual particles. While typical spherical particles in the electrode are shown in blue, particles with already strong fragmentation inside are turquoise and the two red particles display a much weaker absorption and thus an untypical material composition. The picture b) shows a single sectional view with scale bars for size allocation. Due to the wide field of view it is possible to examine a large number of particles and the space between them in one measurement. Images c-e) show enlarged sections of exemplary chosen individual spherical particles in varying degrees of aging. c) is in good condition as the majority of the sample, d) shows already medium fissure as the first consequence of stress, and e) shows an already heavily broken particle. These cracks are a clear sign of the structural degradation of a battery and occur as a result of mechanical or chemomechanical stress during the charging and discharging process. The desirable rapid charging operations in particular accelerate this process.[25] The detection and tracing of these cracks is an important approach in battery research in order to obtain information about the ageing behavior and to develop long-lasting battery types. Using nano CT images, these micro-cracks can be detected in a 3D volume and their location and distribution can be analyzed.

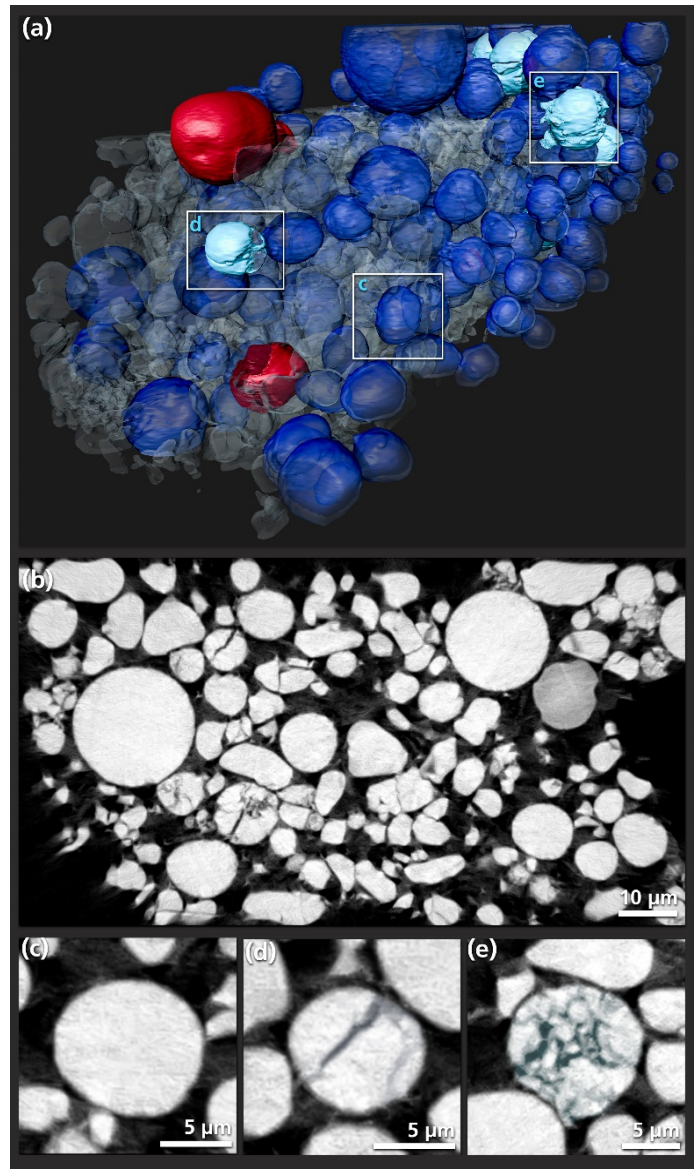

**Figure 7** Reconstruction of the high-resolution nano CT measurement of the electrode of a commercial lithium pouch-cell with 105 nm sampling. (a) 3D rendering of the sample. Typical spherical particles in the electrode are shown in blue, spherical particles with already strong fragmentation inside are turquoise, the two red particles have a much weaker absorption and seem to have a different material composition. (b) unrendered sectional image from the reconstructed volume to get an overview of the ageing state. (c) – (e) enlarged sections of individual spherical particles. (c) in good condition as the majority of the sample. (d) medium fissure as the first consequence of mechanical or chemomechanical stress. (e) already heavily broken particle.

**4.4 Functional Materials Development**

The results of measuring the silicone polymer composite is shown in Figure 8. The picture in a) shows the 3D-rendered volume of the segmented metal particles from the polymer composite. As stated earlier, it is necessary for the further development of this material class to obtain information on how the flake-shaped filler particles are distributed in the polymer matrix. In particular, the alignment of the particles to each other, directional effects from the production or agglomeration of the flakes have to be determined. It should be possible to make statements on the relative relationships of individual particles as well as to depict a sufficiently high number of particles to enable statistical evaluation and find percolation paths. In order to better access these properties visually, a volume section is shown in b). Image c) shows a sectional view of the sample. The sample has previously been subjected to a number of load cycles which have led to a clearly visible agglomeration and local densification of the particles. Furthermore, an enlarged section is shown in d), where the preferred arrangement in stacks and orientation in horizontal direction is therefore visible.

For the further improvement of the composite materials, it is necessary to achieve and prove a strain-stable and fatigue-resistant percolation of the particles. By visualizing a representative volume, nano CT analysis is a valuable tool for the further development of the material class of particle-filled polymer composites.

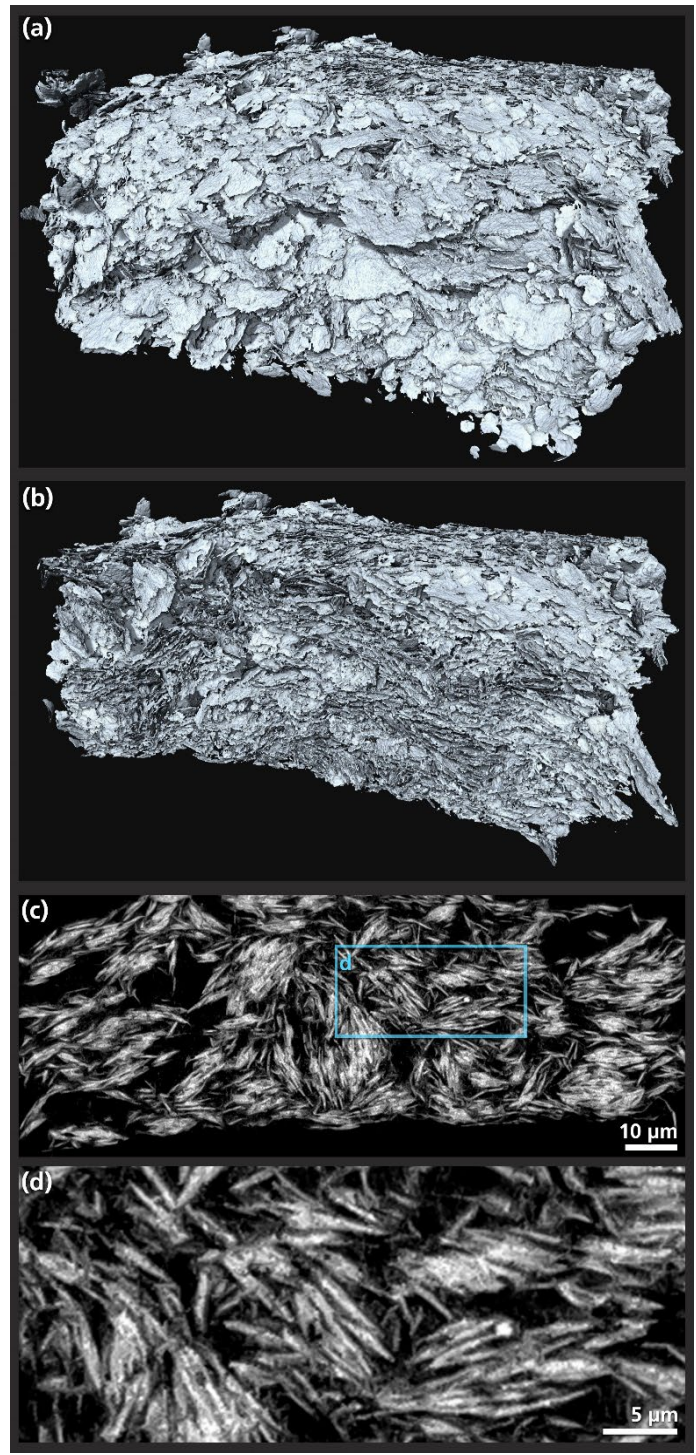

**Figure 8** Reconstruction of the nano CT scan of the functional material composite. (a) 3D rendering of the whole reconstructed sample consisting of a flat cuboid with an edge length of about 100 μm. The rendering focuses on the metallic filler, whose orientation is essential for the function of the material. (b) sliced sample from the above rendering in which the self-organized alignment of the filler can be seen. (c) sectional image from the reconstruction, demonstrating that the particles have aggregated and locally compressed during the cyclic loading. (d) enlarged detail of the sectional image.

**5 Conclusion**

In the present manuscript, a laboratory nano CT system for three-dimensional material analysis based on lens-free X-ray projection magnification with a 2D resolution down to 150 nm has been presented. The validation of the resolution based on the FSC confirmed that a high resolution of about 170 nm - 185 nm has been maintained even in the 3D application. Additionally, our analysis of the contrast in differently sized and presumably periodic structures in an SD card measurement confirms the shape of the system's 2D MTF. This is remarkable and shows the achieved performance of the overall system, since the 3D reconstruction is subjected by more factors potentially influencing the MTF in the 3D reconstruction than in the 2D projection, for example like movement corrections.

Overall, we have demonstrated on the basis of three samples from contemporary fields of research the suitability of the presented system and the potential merit of the laboratory based nano CT analysis for materials research and development. We are capable of visualizing the 100 - 200 nm diameter through silicon vias and examine them for defects like misalignments or insufficient metallization. Therefore, the technology presented here will be a valuable tool for semiconductor research. In the high energy lithium battery electrode, evidence of the battery's performance level was found in the form of chemo-mechanically induced micro-cracks within the electrode particles. Especially in this case, our large field of view enables statistical analysis. Finally, we have demonstrated that for the conductive material composite, despite the highly absorbent material of the filler, the position and orientation of the particles can already be determined visually.

**Acknowledgments**

We would like to thank Dr. Jeffrey Gambino for his expertise and advice on the interpretation of the shown semiconductor structures.

---

[i] The FSC analysis and the calculation of the half-bit criterion were performed in Imagic FSC.[43]

[ii] All graphs shown were plotted in OriginPro 2021 - Origin Lab.

[iii] All sectional images shown were exported with ImageJ Fiji. Also in this software, the local contrast was read out as a line profile of the gray values over the selected structure.[44]

[iv] All 3D visualizations shown in this study were done in Avizo 9.0 – Thermo Fisher Scientific